\documentclass[prl,superscriptaddress,showpacs,preprintnumbers,amsmath,amssymb,twocolumn]{revtex4}
\usepackage{soul}
\usepackage{dcolumn}
\usepackage{bm}
\usepackage{subfigure}
\usepackage{bbding}
\usepackage{graphicx}

\usepackage{amssymb,amsmath}

\begin{document}
\title{Resonant metallic states in driven quasiperiodic lattices }
\author{L. Morales-Molina}
\affiliation{Departamento de F\'isica, Facultad de F\'isica,
 Pontificia Universidad Cat\'olica de Chile, Casilla 306, Santiago 22, Chile
}
\author{E. Doerner}
\affiliation{Departamento de F\'isica, Facultad de F\'isica,
 Pontificia Universidad Cat\'olica de Chile, Casilla 306, Santiago 22, Chile
}
\author{C. Danieli}
\affiliation{New Zealand Institute for Advanced Study, Centre for Theoretical Chemistry and Physics, Massey University,  0745 Auckland, New Zealand}
\author{S. Flach}
\affiliation{New Zealand Institute for Advanced Study, Centre for Theoretical Chemistry and Physics, Massey University,  0745 Auckland, New Zealand}

\begin{abstract}
We consider a quasiperiodic Aubry-Andre (AA) model and add a weak time-periodic and spatially quasiperiodic perturbation. 
The undriven AA model is chosen to be well in the insulating regime. 
The spatial quasiperiodic perturbation extends the model into two dimensions in reciprocal space. For a spatial resonance which
reduces the reciprocal space dynamics to an effective one-dimensional two-leg ladder case,
the ac perturbation resonantly couples certain groups of localized eigenstates of the undriven AA model and turns them into extended metallic ones.
Slight detuning of the spatial and temporal frequencies off resonance returns these states into localized ones.
We analyze the details of the resonant metallic eigenstates using Floquet representations. In particular, we find that their size grows
linearly with the system size. Initial wave packets overlap with resonant metallic eigenstates and lead to ballistic spreading.
\end{abstract}

\maketitle
{\sl Introduction --}
More than 50 years since the publication  of Anderson's seminal paper on wavefunction localization in disordered lattices \cite{anderson}, 
and more than 30 years since the observation of wavefunction localization in quasiperiodic lattices by Aubry and Andre \cite{AAmodel},
the field still records growing interest and activity. 
This is to a large extent
due to the rapid development experienced in the field of cold atoms in optical lattices \cite{bloch}, and of light propagation in structured media \cite{lederer}.
Controllability of the system dimensions, tunability of the interaction among atoms and nonlinearity for propagating light beams, and the 
implementation of designed disorder potentials opens avenues for new challenges. In particular, this has allowed to study i) Anderson localization for  non-interacting atoms \cite{Billy} and
light \cite{AAlight}, and ii) Aubry-Andre (AA) localization for non-interacting atoms \cite{roati} and
for light \cite{lahini09}.

Unlike the one-dimensional Anderson model, where all eigenstates are localized irrespective of the strength of the disorder potential, 
the AA model of a one-dimensional quasiperiodic lattice allows for the existence of extended states as well \cite{michele}.  
The metal-insulator transition (MIT) separates metallic (extended) states from the insulating (localized) states and is controlled by the depth of the quasiperiodic potential.  
Notably the MIT for the AA model is energy-independent due to a duality symmetry \cite{AAmodel}. 
Extensions which include driven ac forces allow additional control over the MIT \cite{Holthaus}.

Here we address the question whether a weak but resonant space-quasiperiodic and time-periodic perturbation can destroy AA localization. Early studies for Anderson localization with uncorrelated random potentials
show that the absence of correlations allows for a finite increase of the localization length in the presence of time-periodic ac perturbations, but keeps a finite upper bound on it \cite{yamada,Rmolina}.
Despite a number of further publications on driven Harper models  \cite{Iomin,dima} and versions of driven AA models \cite{kolovsky}, the results are basically soft modifications of the properties
of the unperturbed eigenstates in the presence of driving.
We consider a time periodic moving lattice which mimics the effect of a driving force. The amplitude of the moving lattice is taken small, but its oscillation frequency is chosen such that it resonantly couples localized states from distinct bands of the undriven AA system. 
At the same time the additional spatial quasiperiodic perturbation effectively increases the dimensionality from $d=1$ to $d=2$ in reciprocal space. At a spatial resonance this dynamics is reduced to an effective one-dimensional ladder topology.
As a result of this resonant coupling delocalization takes place. 
We use Floquet representation to extract the eigenstates of the driven system. We find that at certain resonant values of the driving frequency, groups of eigenstates completely delocalize.
We study the properties of these resonant metallic eigenstates, and perform wave packet evolution tests to show that off resonance AA localization holds, but on resonance ballistic transport
is obtained.

{\sl The Model --}
Ultra-cold atoms in optical lattices offer an ideal benchmark for the study of Anderson localization since disordered lattices are experimentally feasible to build. For instance, a quasiperiodic lattice
potential can be created by a bichromatic lattice \cite{bichro} of the form
\begin{equation}\label{Eq:Potencial}
U(x)=U_{1}\cos(k_{1}x)+U_{2}\cos(k_2x+\phi),
\end{equation}
where $\phi$ is a constant phase introduced to shift the two lattices relative to each other. Here $k_{1}$ and $k_2$ are the wavevectors and $U_{1}$ and $U_{2}$ are the amplitudes of the two 
lattice potentials. $E_{r}=(\hbar k_L)^2/2M$ is the recoil energy of the first lattice and $M$ is the mass of the atoms. 

Within the tight-binding approximation, the quantum dynamics for a particle moving in a quasiperiodic potential  Eq.\ref{Eq:Potencial} can be described 
by the Harper \cite{harper} or the Aubry-Andr\'e  (AA) model \cite{AAmodel}. Details of the derivation of the AA model can be found for instance in Ref.\cite{giamarchi}. 
Here we consider the driven AA model 
\begin{align}
 \label{Eq:shrod}
 i \frac{d\psi_{n}}{dt}&= 
\psi_{n+1}+\psi_{n-1}+ V_{1}\cos(2\pi \alpha n+\delta)\psi_{n}
\nonumber
\\
&
+V_{2}\cos(2\pi \beta n +\Omega t)\psi_{n}\;.
\end{align}
The first line in Eq.(\ref{Eq:shrod}) corresponds to the undriven AA model in dimensionless units. Here $\psi_n$ is the complex wave function amplitude at lattice site $n$, $V_1$ is the 
strength of the quasiperiodic potential, $\alpha$ is an irrational number setting the spatial period $1/\alpha$ which is incommensurate with the lattice spacing $\Delta n \equiv 1$ (here $\alpha$ will be chosen as  $\alpha=\sqrt{3}-1\approx 0.732$),
and $\delta$ is a relative phase.
The second line in Eq.(\ref{Eq:shrod}) represents the time-periodic perturbation potential with amplitude $V_2$, spatial period $1/\beta$ and the ac driving frequency $\Omega$.
Note that the equations are invariant under shifts of $\alpha$ or $\beta$ by any integer, therefore their irreducible space is confined
to the unit interval.
Despite the time-dependent perturbation the above equations enjoy a generalized symmetry $\beta \rightarrow -\beta$, $t\rightarrow -t$ and
$\psi \rightarrow \psi^*$. 

The perturbation potential can be generated experimentally using a similar function as in Eq.(\ref{Eq:Potencial}),  where the overall phase is now a function of time. 
\begin{figure}[h!]
\includegraphics[width=0.9\columnwidth]{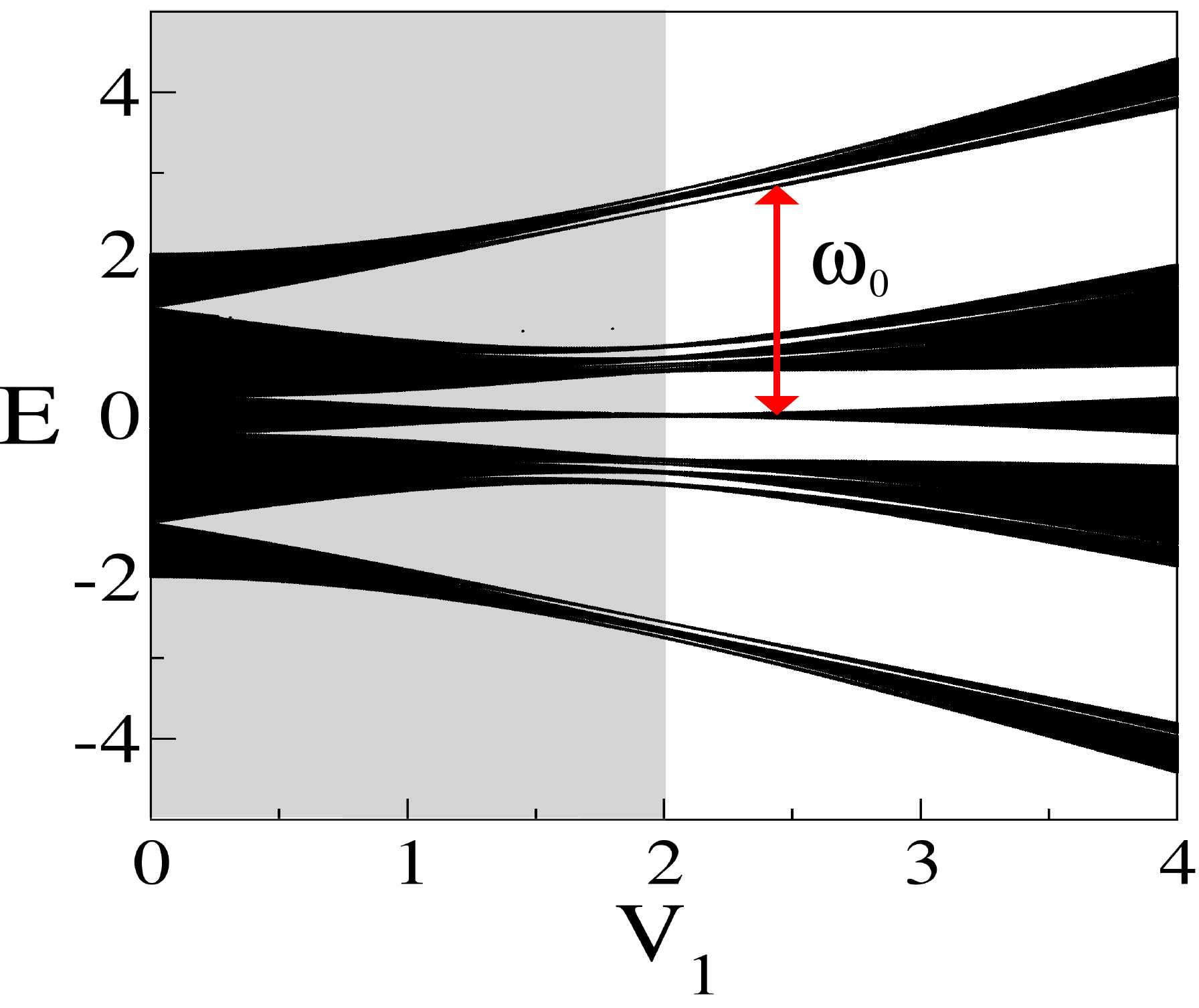}
\caption{Energy spectrum $E$ vs. $V_{1}$ of the undriven Aubry-Andre model ($V_{2}=0$) with $\alpha=\sqrt{3}-1$.  The shadowed region indicates the metallic phase for $V_{1}<2$. 
The arrow indicates the separation between edge and center bands for $V_{1}=2.5$ within the insulating region.}
\label{fig1}
\end{figure}

In the absence of ac driving, $V_{2}=0$, the system (\ref{Eq:shrod}) is self-dual \cite{AAmodel} and it is known to possess a MIT 
at the critical depth of the potential $V_1=2$ \cite{AAmodel}. This is usually shown through a duality transformation. We generalize this transformation to the case of our driven model.  
In analogy to the procedure in \cite{AAmodel} we define
\begin{equation}
\psi_n \equiv {\rm e}^{i 2 \pi K n} \sum_{l,m} f_{l,m}{\rm e}^{i \left[ (2 \pi\alpha n + \delta)l + (2 \pi\beta n + \delta)m \right] }\;.
\label{2dduality}
\end{equation}
Here $f_{lm}$ are the complex wave function amplitudes in reciprocal (spatial Fourier) space with the Bloch wave number $K$. The indices $l,m$ can take all integer values. They correspond to the two lattice frequencies $2\pi \alpha$ and $2\pi \beta$ of the model. So far we consider
the case when the ratio $\alpha / \beta$ is irrational as well.  Then the equations (\ref{Eq:shrod}) turn into a two-dimensional lattice 
problem in reciprocal space:
\begin{align}
i\frac{d f_{l,m}}{dt} &=  2 \cos \left[ 2\pi (\alpha l + \beta m + K) \right] f_{l,m}  
\nonumber
\\
& + \frac{V_1}{2} ( f_{l+1,m} + f_{l-1,m}) 
\nonumber
\\
& + \frac{V_2}{2} \left( {\rm e}^{i 2 \pi g(t)} f_{l,m-1} + 
 {\rm e}^{-i 2 \pi g(t)} f_{l,m+1} \right)
\label{2dfourier}
\end{align}
where the function
\begin{equation}
g(t) = \Omega t - \delta
\label{phase}
\end{equation}
takes care of the time-dependent relative phase shifts of the two quasiperiodic potentials. The structure of the third line in Eq.(\ref{2dfourier}) suggests that the second perturbation potential acts as a magnetic field. However it is not hard to see that the total flux per plaquette in the
two-dimensional reciprocal lattice $\{l,m\}$ is zero.

For $V_2=0$ the two-dimensional lattice equations (\ref{2dfourier}) decouple into independent and equivalent one-dimensional ones.
These chains are identical with the unperturbed Aubry Andre case in real space (\ref{Eq:shrod}) and exhibit a MIT at the critical value $V_1=2$.
With the ansatz $\psi_n(t) = {\rm e}^{iEt} \phi_n$ we solve the corresponding eigenvalue problem for the eigenenergies $E$.
Fig.\ref{fig1} shows the spectrum of energy $E$ as function of $V_{1}$. 
Well defined gaps appear in the spectrum between different eigenenergies, due to its topological structure (for all $V_1\neq0$ the spectrum is a Cantor set \cite{Av1}). However, the nature of the corresponding eigenmodes depends on $V_1$, since for $V_{1}>2$ all the eigenstates are localized whereas for $V_{1}<2$ they are extended. This means that any wavepacket  
for $V_{1}>2$ will remain localized.

Spatial resonances happen for $V_2 \neq 0$ but with a commensurate ratio of the two spatial frequencies $\alpha / \beta = p/q$ with $\{p,q\}$ integers. Then the transformation (\ref{2dduality}) extends only over a $q$ independent integers $m=0,1,2,...,q-1$, turning the two-dimensional
lattice (\ref{2dfourier}) into a one-dimensional ladder with $q$ legs. The boundary conditions between the 1st and the $q$th leg are defined
by the second integer $p$.

{\sl Driven Lattice: Resonances --} 
Let us add a weak perturbation potential $V_2 =0.25$ for the case
$V_{1}=2.5$, where the eigenstates of the undriven system are localized.
For the undriven system, the eigenenergies of these states form bands which are separated by gaps (Fig.\ref{fig1}). 
We hunt for a resonant coupling of these states through the ac driving, therefore the driving frequency $\Omega$ should take values
of the order of the gaps. 
  
\begin{figure}[h!]
\includegraphics[width=0.9\columnwidth]{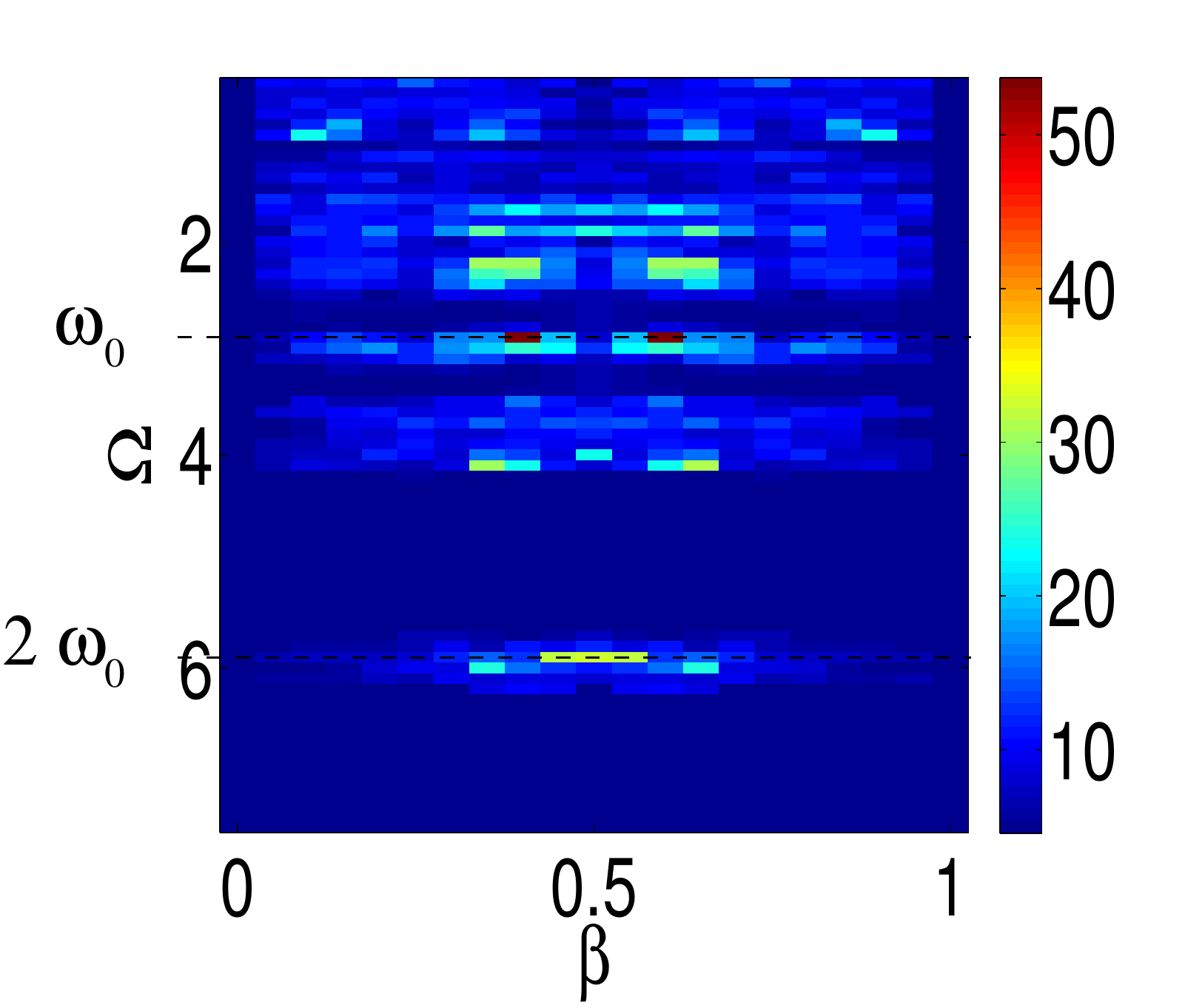}
\includegraphics[width=0.9\columnwidth]{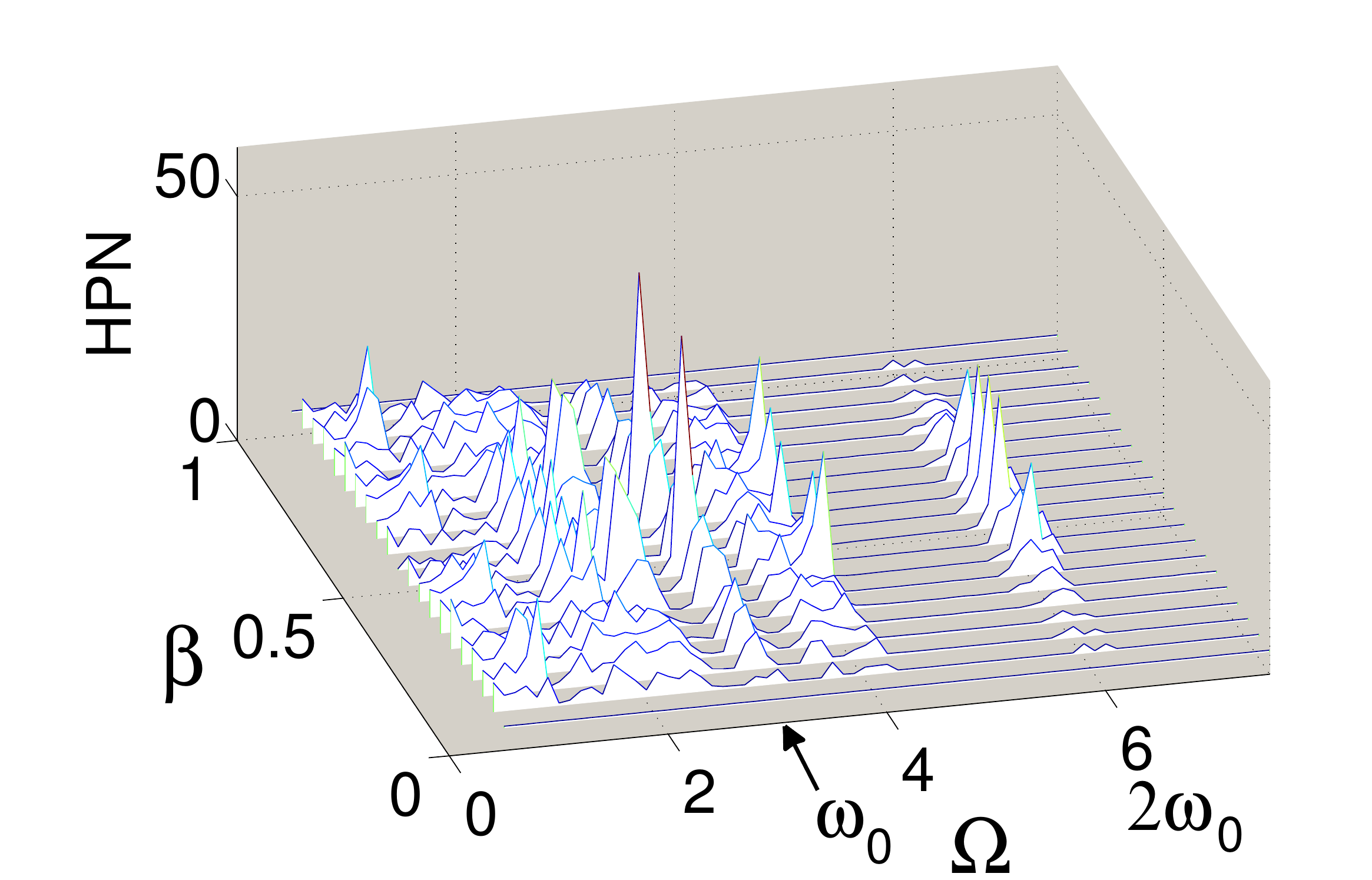}
\caption{(Color online) Upper plot: Density plot of highest participation number $HPN$ vs $\Omega$ and $\beta$. Dashed lines indicates the resonances $\Omega=\omega_{0}=2.9$ and $\Omega=2\omega_{0}=5.8$. Bottom plot: same in a three dimensional plot, with good visibility of the
strong resonance at $\beta\approx 0.37$. Here $V_1=2.5$, $V_2=0.25$, $\alpha=\sqrt{3}-1$.
}
\label{fig2}
\end{figure}

Due to the time-periodic forcing, it is convenient to use the Floquet formalism.
The  wavefunction is expanded in terms of Floquet states  
$u_{n}(t)=e^{i\epsilon t}\phi_{n}(t)$, with $\phi(t+T)_{n}=\phi_{n}(t)$, where $T=2\pi/\Omega$ (see also \cite{denisov} for details).
 To determine the degree of (de)localization of Floquet states, we compute the participation number \cite{local,localength}
\begin{equation}\label{Eq:PR}
 PN=\frac{\sum_{n}|u_{n}|^2}{\sum_{n}|u_{n}|^4}.
\end{equation}
with the lower bound $PN=1$ which accounts for maximal localization, and the upper bound $PN=N$ corresponding to a perfectly extended state
(here $N$ is the number of driven lattice sites).  For each parameter set $\{\beta,\Omega\}$ we find all Floquet eigenstates, compute $PN$ for each of them,
and identify the highest value $HPN$.   
Fig.\ref{fig2} shows the intensity plot  of $HPN$ as a function of $\beta$ and $\Omega$. 
While even the states with largest spatial extend stay very localized for almost all parameter values, we find a strong resonant spot at $\beta \approx 0.37$ and $\Omega\approx 2.9$, and its symmetry related partner point at $\beta\approx 0.63$.  Weaker spots are observed at
$\Omega\approx 2.4,4,5.8$.

\begin{figure}[h!]
\includegraphics[width=0.9\columnwidth]{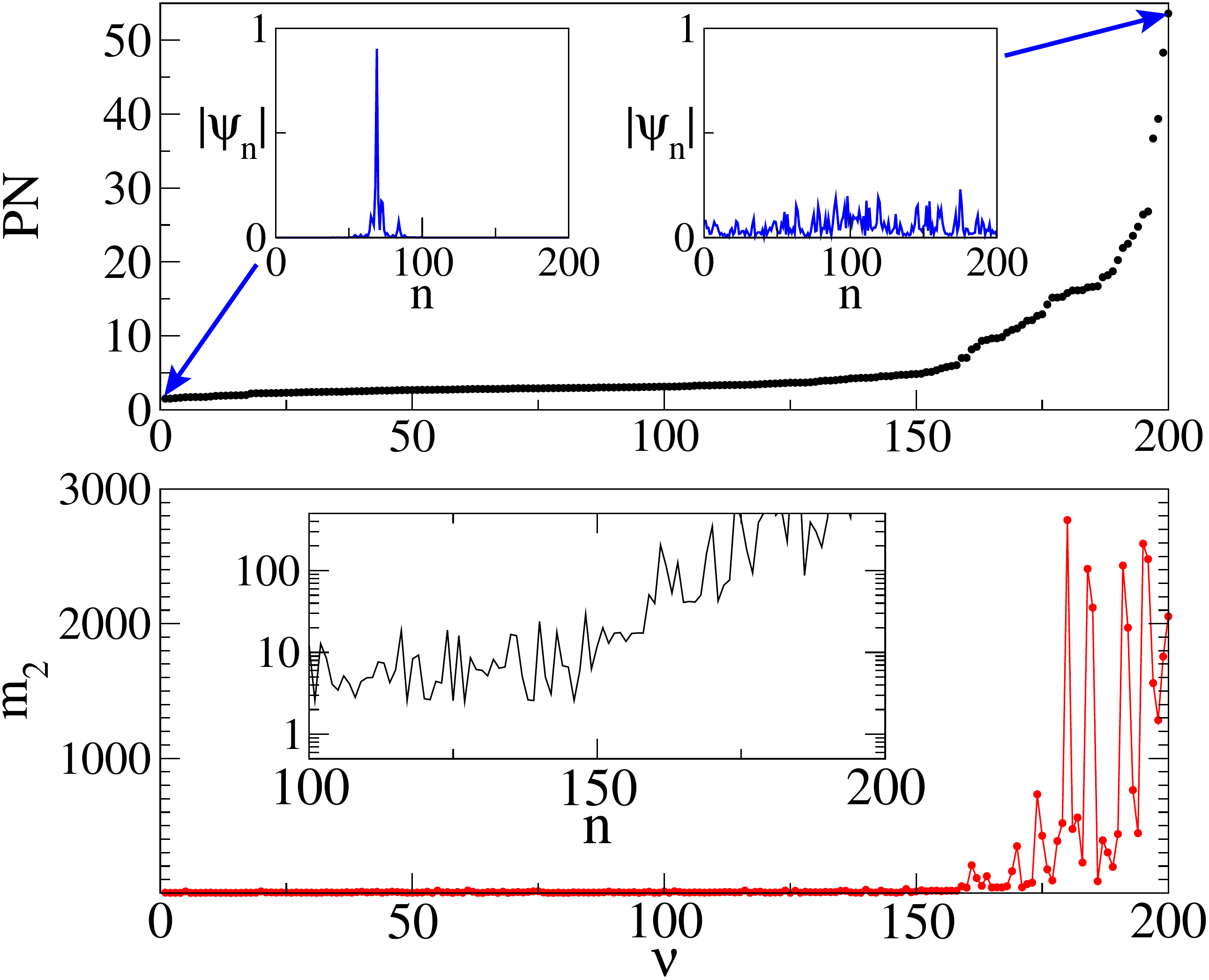}
\caption{(Color online) Upper panel: Ordered values of $PN$ of the Floquet states vs  Floquet state number $\nu$. 
The insets show the absolute value of the wave function versus lattice site $n$ for the state with smallest and largest $PN$.
Lower panel: Same for the second moment 
$m_2$, with the sorting index from the upper panel. Inset shows enlargement of the region around the number 150 for a logarithmic scale. 
Here $V_2=V_1/10=0.25$,  $\alpha=\sqrt{3}-1$, $\beta=0.37$, $\Omega=2.9$.
} 
\label{fig3}
\end{figure}

At the main resonance, $\beta$ corresponds approximately to $\alpha / 2$. Therefore the two-dimensional lattice in reciprocal space (\ref{2dfourier}) is reduced to 
a two-leg ladder. The resonance frequency $\Omega = 2.9$ fits very well the distance between the edge and center bands in Fig.\ref{fig1}.
To make sure that the observed resonance is indeed due to extended metallic states, we plot in Fig.\ref{fig3} the participation numbers $PN$ for
all Floquet eigenstates of a system with size $N=200$ sorted according to their increasing value $\nu$. In addition we also plot their second moments
\begin{equation}\label{Eq:sigma}
 m_2=\sum_{n}(n-\langle n_0\rangle)^2|u_{n}|^2,
\end{equation}
where $\langle n_{0}\rangle= \sum_n n |u_n|^2$ is the first moment of the wavepacket. 
Note that we do not resort the second moments, but plot them using the sorting principle for the participation numbers. We find that 
states with increasing size are detected using both measures. In the inset of the upper panel in Fig.\ref{fig3} 
we show the wave function of the state with the smallest $PN$ and the largest $HPN$. While the former one is strongly localized, the
state with $HPN$ is clearly delocalized over the whole system.

\begin{figure}[h!]
\includegraphics[width=0.9\columnwidth]{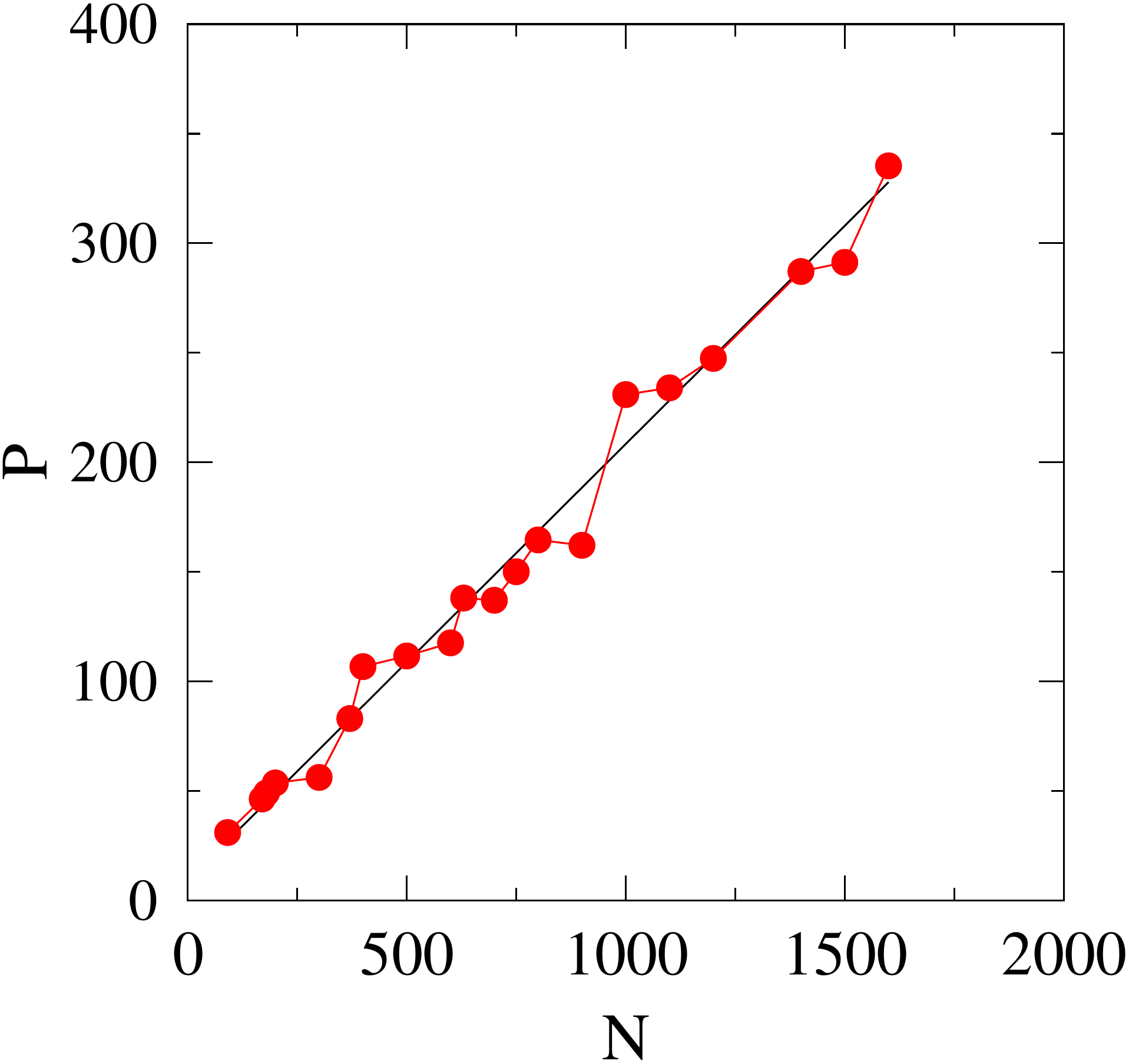}
\caption{(Color online)  Highest participation number $HPN$  as a function of the system size $N$ at resonance $\beta=0.37$, $\Omega=2.9$.
Other parameters $V_1=2.5$, $V_2=0.25$, $\alpha=\sqrt{3}-1$.
} 
\label{fig4}
\end{figure}

If the observed resonance is leading to complete delocalization in an infinite system, the $HPN$ should scale linearly with the system size $N$ of a finite lattice. We test this prediction and show in Fig.\ref{fig4} that it is indeed correct. Therefore the observed resonance generates
metallic eigenstates.
 
\begin{figure}[h!]
\includegraphics[width=0.9\columnwidth]{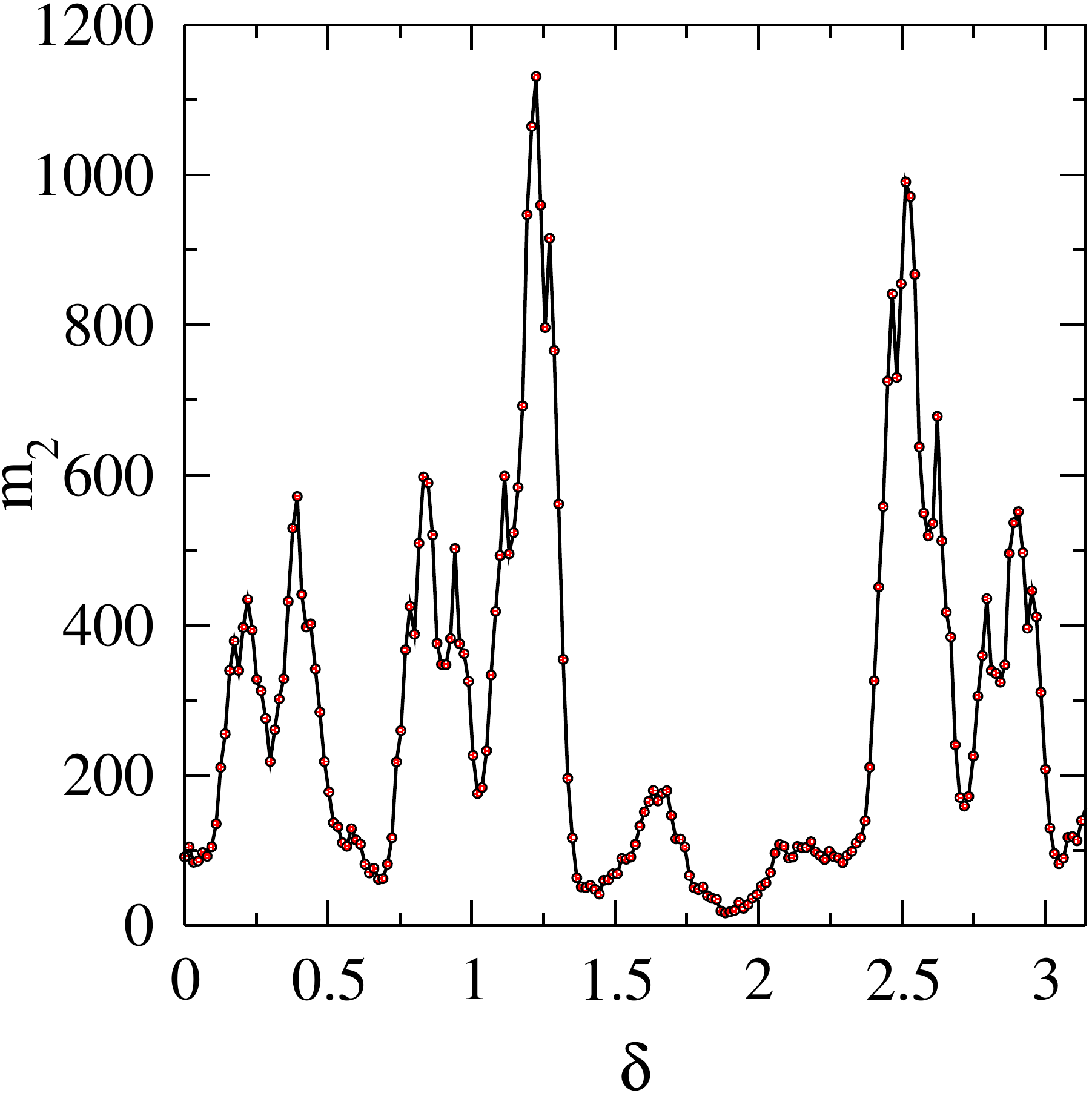}
\caption{Second moment $m_2$ of $|\psi_n(t=10^3)|^2$ as a function of $\delta$  for a single site excitation 
at resonance $\beta=0.37$, $\Omega=2.9$.
Other parameters $V_1=2.5$, $V_2=0.25$, $\alpha=\sqrt{3}-1$, $N=2^8$.
\label{fig5}
}
\end{figure}

\begin{figure}[h!]
\includegraphics[width=0.9\columnwidth]{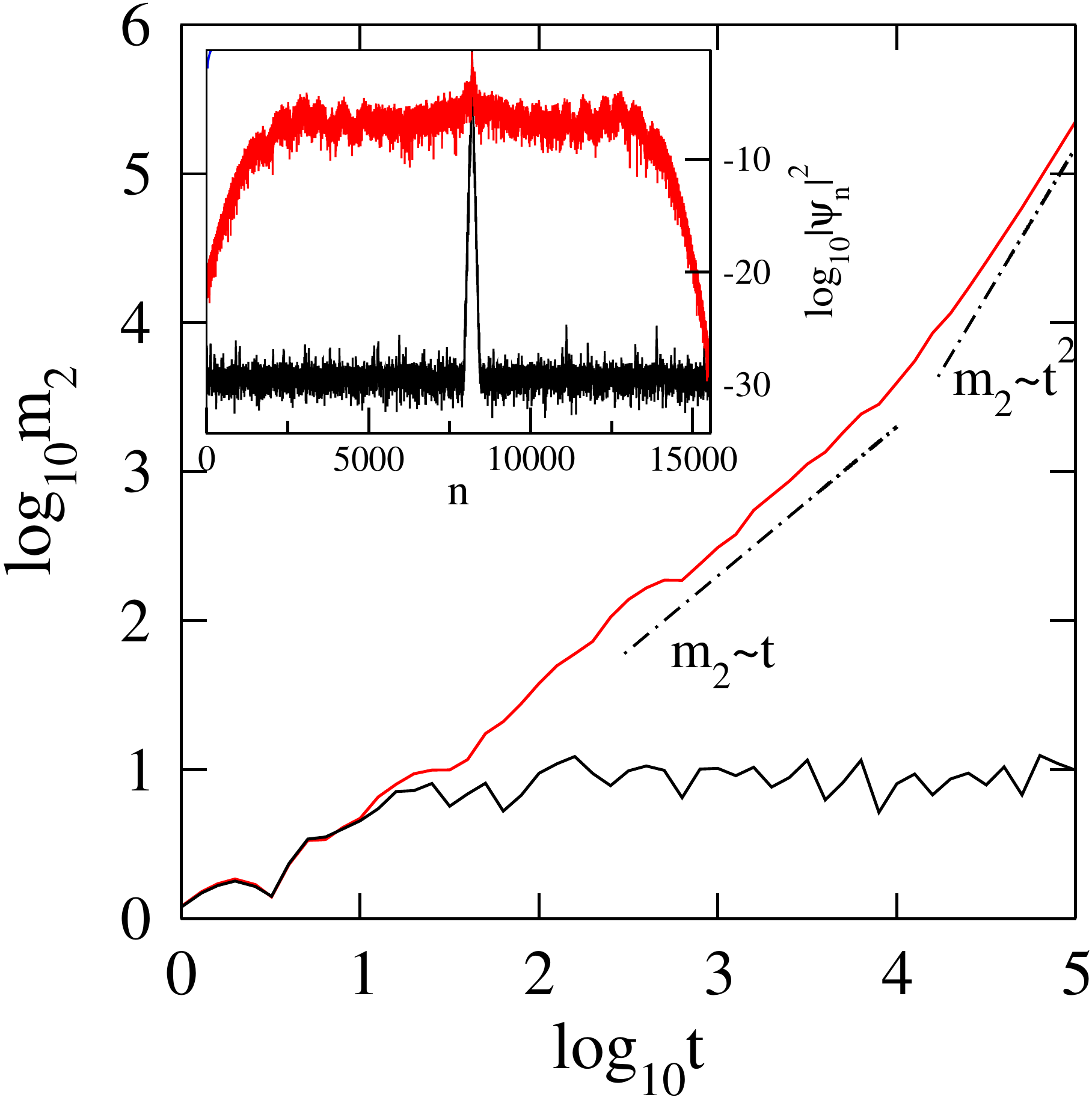}
\caption{(Color online) $m_2$ as a function of time. Red line: $\Omega=2.9$ (on resonance). Black line:  $\Omega=2.6$ (off resonance). In both cases $\beta=0.37$. Dashed-dotted lines indicate normal diffusion $m_2\sim t$ and ballistic spreading $m_2\sim t^2$.
Other parameters $V_1=2.5$, $V_2=0.25$, $\alpha=\sqrt{3}-1$, $N=2^{14}$. 
Inset: density distribution of the wavefunction at the final time of integration. Red extended curve - $\Omega=2.9$ (on resonance), 
black localized curve:  $\Omega=2.6$ (off resonance).
}
\label{fig8}
\end{figure}

{\sl Wave Packet Spreading --}
Another consequence of the generation of delocalized metallic states is that an initially localized wave packet will spread, if at least
a part of it has nonzero overlap with the metallic states. We perform an integration of the time-dependent equation 
(\ref{Eq:shrod}) using a modified SBAB$_2$ symplectic integrator \cite{symplectic} with initial condition $\psi_{N/2}(t=0)=1$ and
$\psi_{n \neq N/2}(t=0)=0$. For the Floquet eigenvalue problem, the eigenvalues and eigenvectors are insensitive to the value of the relative 
phase $\delta$ (see Eq.(\ref{Eq:shrod})), which essentially shifts the potential relative to the lattice. The overlap of a given initial
state with Floquet eigenstates however does depend in general on the location of the initial state, or equivalently on the value
of $\delta$. We perform short runs until time $t=10^3$ with system size $N=2^8$. 
and measure the second moment $m_2$ as a function of $\delta$ (Fig.\ref{fig5}).
As expected we find strong fluctuations due to varying overlap. We choose $\delta \approx 1.225$ and perform long-time runs up to $t=10^5$. We plot the time-dependence of the second moment $m_2(t)$ on logarithmic scales
in Fig.\ref{fig8}.
The red line in Fig.\ref{fig8} shows indeed that the second moment increases with time, and turns into a ballistic $t^2$ law after a transient
$t\approx 10^4$. The inset shows the density distribution at the final integration time which is wide spread over the lattice.
A test run off resonance with $\Omega=2.6$ shows no signature of spreading, a small second moment which is constant on average, and a highly
localized wave function density at the final integration time.

\begin{figure}[h!]
\includegraphics[width=0.9\columnwidth]{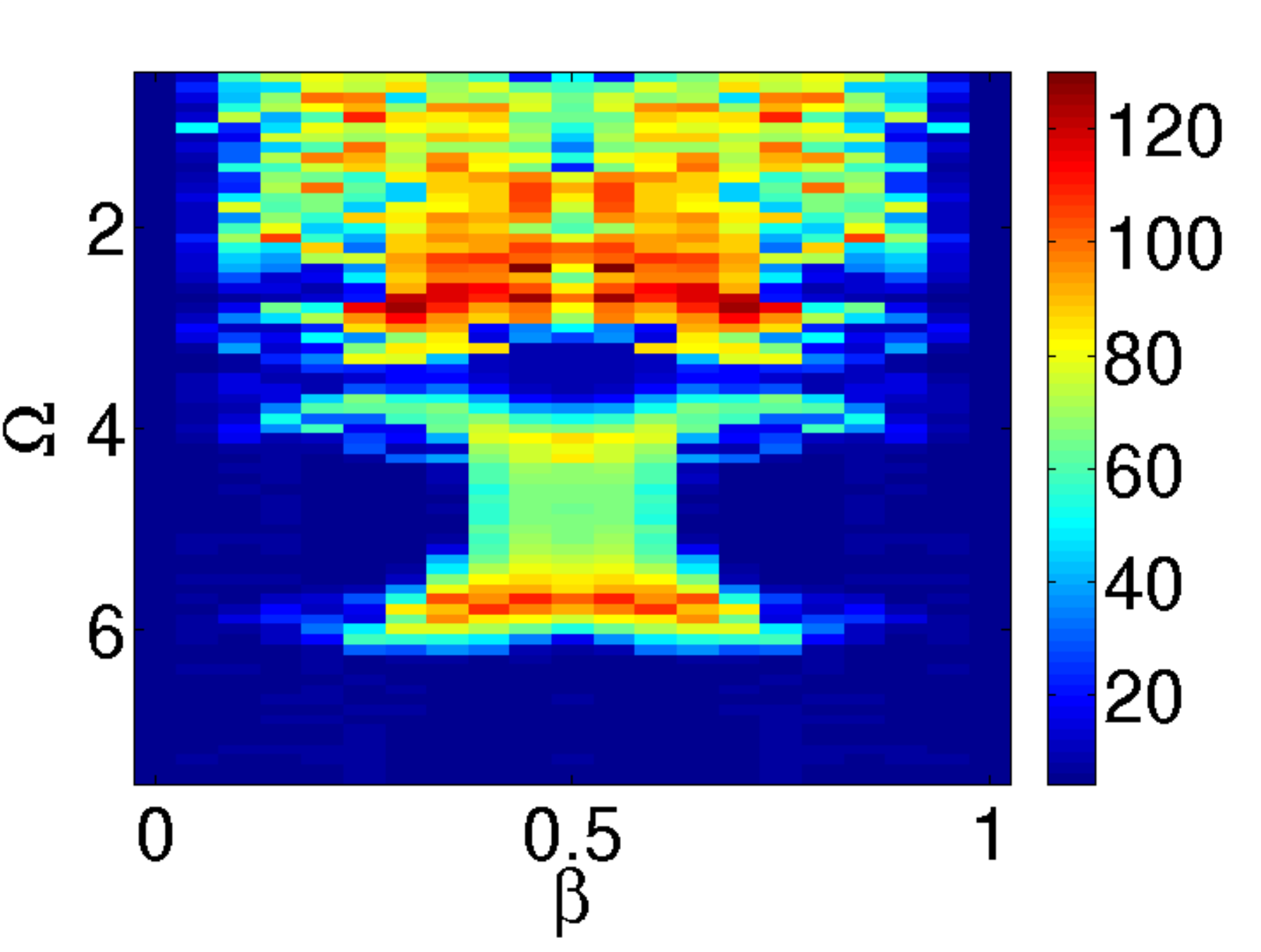}
\caption{(Color online) Density plot of the highest participation number $HPN$ vs $\Omega$ and $\beta$. Here $V_1=2.5$, $V_2=0.5$, $\alpha=\sqrt{3}-1$.}
\label{fig9}
\end{figure}

In the presence of moderate and strong driving the observed resonances are expected to broaden. We test this prediction repeating the calculation
in Fig.\ref{fig2} while doubling the strength of the driving potential $V_2=0.5$. The outcome is shown in Fig.\ref{fig9}.
We indeed confirm that the metallic regime widens substantially.

{\sl Concluding Remarks --}
We have studied the Aubry-Andre system in the presence of a weak driving force. The driving force is introduced upon superimposing a moving time periodic lattice  over a stationary quasiperiodic potential. 
The undriven AA model is chosen to be well in the insulating regime. 
The spatial quasiperiodic perturbation extends the model into two dimensions in reciprocal space. For a spatial resonance which
reduces the reciprocal space dynamics to an effective one-dimensional two-leg ladder case,
the ac perturbation resonantly couples groups of localized eigenstates of the undriven AA model and turns them into extended metallic ones.
Slight detuning of the spatial and temporal frequencies off resonance returns these states into the family of localized ones.
Initial wave packets overlap with resonant metallic eigenstates and lead to ballistic spreading.
Therefore the proposed ac perturbation can be used as a simple and elegant method to control the degree of localization
of wave packets in quasiperiodic lattices.
\\
\\
Acknowledgments.
\\
We thank Xiaoquan Yu for help with the SBAB$_2$ code. L.M-M acknowledges financial support from the FONDECYT projects no 1110671 and 1130705.

\end{document}